# Electrically Reconfigurable Nonvolatile Metasurface Using Low-Loss Optical Phase Change Material

**Yifei Zhang[1], Clayton Fowler[2], Junhao Liang[1], Bilal Azhar[1], Mikhail Y. Shalaginov[1], Skylar Deckoff-Jones[1], Sensong An[2], Jeffrey B. Chou[3], Christopher M. Roberts[3], Vladimir Liberman[3], Myungkoo Kang[4], Carlos Rίos[1], Kathleen A. Richardson[4], Clara Rivero-Baleine[6], Tian Gu[1,5], Hualiang Zhang[2,\*], and Juejun Hu[1,5,\*]**

[1]*Department of Materials Science & Engineering, Massachusetts Institute of Technology, Cambridge, MA, USA*
[2]*Department of Electrical & Computer Engineering, University of Massachusetts Lowell, Lowell, MA, USA*
[3]*Lincoln Laboratory, Massachusetts Institute of Technology, Lexington, MA, USA*
[4]*The College of Optics & Photonics, Department of Materials Science and Engineering, University of Central Florida, Orlando, FL, USA*
[5]*Materials Research Laboratory, Massachusetts Institute of Technology, Cambridge, MA, USA*
[6]*Missiles and Fire Control, Lockheed Martin Corporation, Orlando, FL, USA*

\**hualiang_zhang@uml.edu, hujuejun@mit.edu*

**Abstract**

Active metasurfaces promise reconfigurable optics with drastically improved compactness, ruggedness, manufacturability, and functionality compared to their traditional bulk counterparts. Optical phase change materials (O-PCMs) offer an appealing material solution for active metasurface devices with their large index contrast and nonvolatile switching characteristics. Here we report what we believe to be the first electrically reconfigurable nonvolatile metasurfaces based on O-PCMs. The O-PCM alloy used in the devices, $Ge_2Sb_2Se_4Te_1$ (GSST), uniquely combines giant non-volatile index modulation capability, broadband low optical loss, and a large reversible switching volume, enabling significantly enhanced light-matter interactions within the active O-PCM medium. Capitalizing on these favorable attributes, we demonstrated continuously tunable active metasurfaces with record half-octave spectral tuning range and large optical contrast of over 400%. We further prototyped a polarization-insensitive phase-gradient metasurface to realize dynamic optical beam steering.

DISTRIBUTION STATEMENT A. Approved for public release.
Distribution is unlimited. This material is based upon work supported by the Under Secretary of Defense for Research and Engineering under Air Force Contract No. FA8702-15-D-0001. Any opinions, findings, conclusions or recommendations expressed in this material are those of the author(s) and do not necessarily reflect the views of the Under Secretary of Defense for Research and Engineering.

## Introduction

Metasurfaces, planar electromagnetic artificial media comprising arrays of subwavelength-scale antennas or "meta-atoms", have been widely regarded as a promising platform enabling optical elements with significant potential Size, Weight, Power, and Cost (SWaP-C) benefits over their conventional bulk counterparts. Active metasurfaces, whose optical responses can be dynamically modulated, further enable exciting opportunities for agile manipulation of light propagation and interaction with matter[1–7]. To date, active tuning of metasurfaces has leveraged mechanical deformation[8–10], electrochemical or chemical reactions[11,12], electro-optic and thermo-optic effects[13–15], as well as phase change media[16–29]. In particular, phase change materials (PCMs) based on chalcogenide alloys are uniquely poised for realizing phase-gradient active metasurfaces, since the giant refractive index contrast imparted by their structural transition facilitates broad optical phase tuning[30]. For instance, we have recently demonstrated that full $2\pi$ phase coverage can be achieved in PCM-based active metasurfaces, enabling binary switching between arbitrary phase profiles[31,32].

To date, active modulation of PCM-based metasurfaces has relied upon thermal annealing[21,28,31,33–35] or optical writing[22–25,36–39]. These modulation schemes, however, necessitate bulky heating furnaces or ultrafast lasers. Electrical switching of PCM, on the other hand, is naturally conducive to compact integration with flat optics and foresees miniaturized, chip-scale reconfigurable optical systems. While electrical addressing of PCM is already a mature technology in phase-change random-access memories (PCRAMs), the much larger device area essential for optical devices – as compared to today's deeply-scaled PCRAMs – places more stringent requirements on switching homogeneity across large areas. This technical challenge is epitomized by the phenomenon of filamentation, which precludes switching by passing current directly through PCM[30]. To address this challenge, we report in this work the first experimental demonstration of large-area, reconfigurable metasurfaces based on electrothermal switching of PCM. Our work advances the state-of-the-art in three important aspects. First, our devices are made of a newly developed PCM, $Ge_2Sb_2Se_4Te_1$ or GSST[40–43]. Compared to the prevailing GST alloys, GSST offers two unique advantages specific to active metasurfaces: its broadband transparency across different structural states mitigates optical losses, and its much larger switching volume allows for optically thick PCM structures to boost light-PCM interactions while maintaining dynamic and fully reversible switching capacity[22,42]. The latter feature underpins the giant optical contrast and tuning range of our devices. Second, we have demonstrated voltage-controlled multi-state tuning of the PCM metasurface covering a record half-octave spectral regime. While multi-state operation has been reported in PCM-integrated waveguide devices[44–47], we show that progressive phase transition in our devices proceeds in a distinctively different, spatially uniform manner facilitating precise meta-atom tuning. Last but not least, we have realized large-area (up to 0.4 mm × 0.4 mm), uniform electrothermal switching of PCMs using geometrically optimized heaters, which heralds new topologically optimized heater designs towards ultimate scaling of active metasurface aperture size.

## Electrothermal switching device platform

Each electrically reconfigurable metasurface device comprises meta-atoms resting on a metal heater, which also acts as a reflector (Fig. 1a). The meta-atoms are patterned in a GSST film and the fabrication process details are described in Methods. The metasurface device arrays were then wire-bonded to a printed circuit board carrier with pluggable ribbon cable connections to facilitate computerized control of the devices (Fig. 1b). In each device, the structural state of the meta-atoms is collectively controlled by electrical pulsing, where a long, low-voltage (500 ms duration, < 12

V) pulse triggers crystallization via Joule heating, whereas a short, high-voltage (5 μs, 20-23 V) pulse re-amorphizes the meta-atoms via a melt-quench process (Fig. 1c).

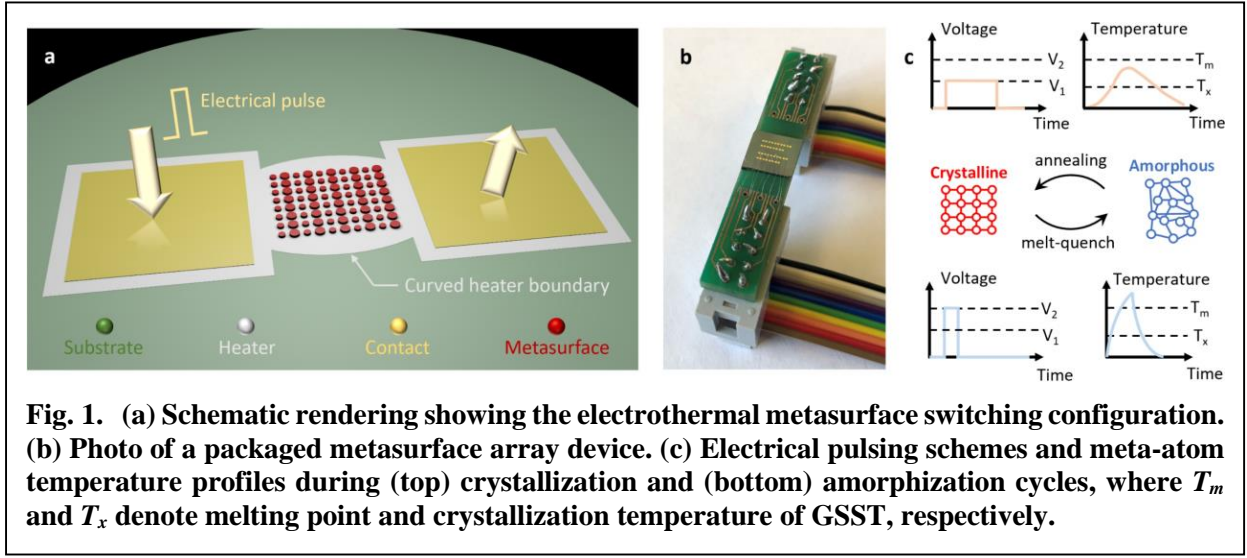

**Fig. 1.** (a) Schematic rendering showing the electrothermal metasurface switching configuration. (b) Photo of a packaged metasurface array device. (c) Electrical pulsing schemes and meta-atom temperature profiles during (top) crystallization and (bottom) amorphization cycles, where $T_m$ and $T_x$ denote melting point and crystallization temperature of GSST, respectively.

PCM metasurfaces with square apertures of varying sizes (from 0.1 mm × 0.1 mm to 0.4 mm × 0.4 mm) were fabricated. Compared to previously reported electrothermal PCM switching devices[42,48,49], the heater unit area reported here is more than two orders of magnitude larger. Mitigation of thermal non-uniformity becomes a critical challenge with enlarged metasurface aperture size, since large temperature excursions 1) preclude precise multi-state tuning due to non-uniform switching during crystallization and 2) compromise device reliability owing to emergence of "hot spots" during amorphization. To illustrate this challenge, Fig. 2a plots the simulated steady-state temperature profile during a crystallization pulse for a classical square-shaped heater within the 0.15-mm device aperture. It is apparent that temperature near the perimeter of the aperture is considerably lower due to more efficient heat dissipation. To counter the issue, we introduced curved heater boundaries (Fig. 1a) to both lower current density in the heater center and move the heater boundary further away from the metasurface aperture. Figure 2b presents the corresponding temperature profile for a heater with optimized boundary curvature (Supplementary Section I) showing significantly improved in-plane uniformity. Based on the optimized heater design, Fig. 2c maps the average heater temperature for different pulse duration and voltage combinations, and the results (550 – 630 K for crystallization and 1020 K for amorphization) are in excellent agreement with our prior data[42,50].

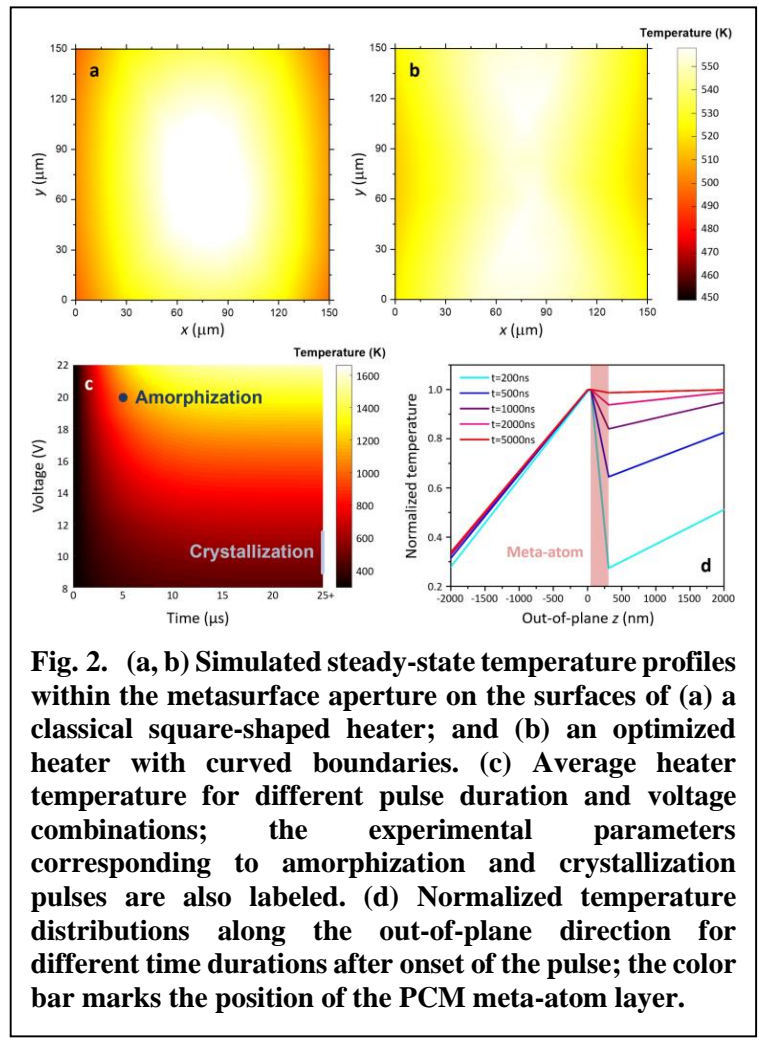

**Fig. 2.** (a, b) Simulated steady-state temperature profiles within the metasurface aperture on the surfaces of (a) a classical square-shaped heater; and (b) an optimized heater with curved boundaries. (c) Average heater temperature for different pulse duration and voltage combinations; the experimental parameters corresponding to amorphization and crystallization pulses are also labeled. (d) Normalized temperature distributions along the out-of-plane direction for different time durations after onset of the pulse; the color bar marks the position of the PCM meta-atom layer.

We further investigated thermal uniformity inside meta-atoms along the out-of-plane (through-thickness) direction. Figure 2d plots the temperature distributions along the out-of-plane direction for different time durations after onset of the electrical pulse, all normalized to the respective peak temperature at the heater surface. Within 5 μs (duration of the amorphization pulse), the temperature distribution has approached the steady state with an average temperature deviation of only 4 K throughout the entire meta-atom. The negligible temperature gradient in the meta-atoms, coupled with the reduced temperature sensitivity of crystallization kinetics in PCMs at temperatures well above glass transition[51], indicate that the transition proceeds progressively and uniformly throughout the meta-atoms during crystallization, leading to a mixture of interspersed crystalline and amorphous phases. This contrasts with intermediate states of PCM elements integrated with waveguide devices, where the crystalline and amorphous phases are spatially localized[45]. The meta-atom tuning mechanism is viable since GSST is classified as a nucleation-dominated PCM[42], such that GSST nanocrystals can precipitate and grow from an amorphous matrix throughout the meta-atom volume. The ability to precisely access different intermediate states with voltage control is critical to enabling multi-state operation of the PCM active metasurface, a key attribute of our strategy.

**Active metasurface characterization**

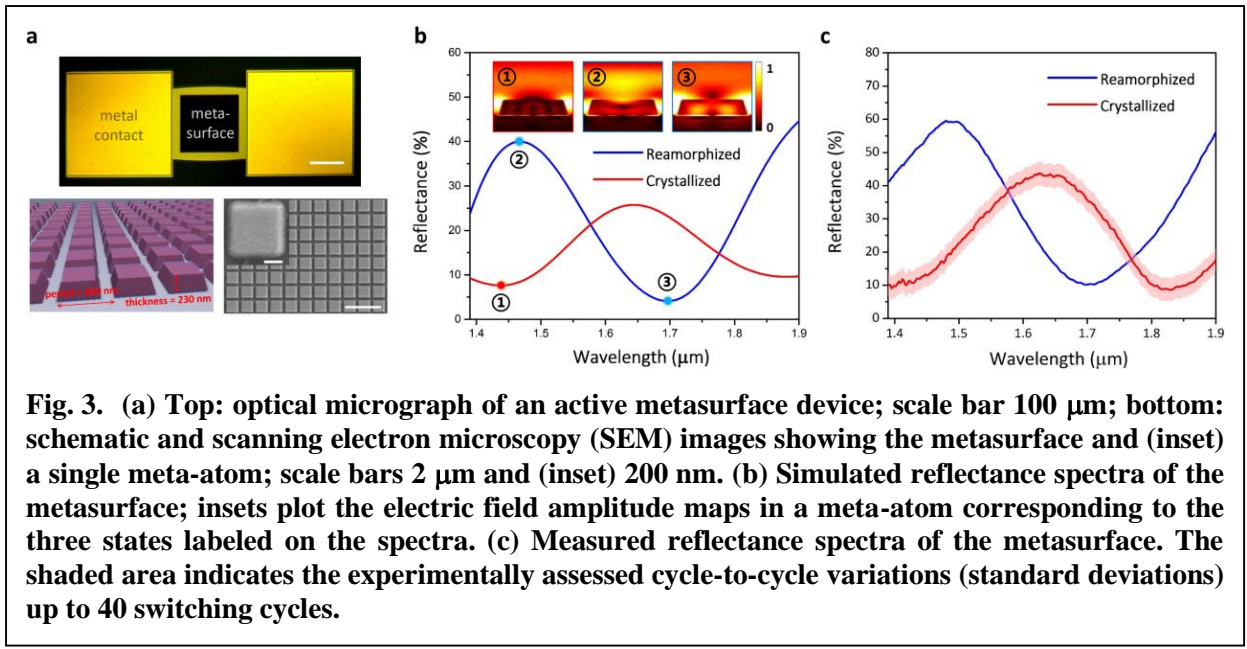

Fig. 3. (a) Top: optical micrograph of an active metasurface device; scale bar 100 μm; bottom: schematic and scanning electron microscopy (SEM) images showing the metasurface and (inset) a single meta-atom; scale bars 2 μm and (inset) 200 nm. (b) Simulated reflectance spectra of the metasurface; insets plot the electric field amplitude maps in a meta-atom corresponding to the three states labeled on the spectra. (c) Measured reflectance spectra of the metasurface. The shaded area indicates the experimentally assessed cycle-to-cycle variations (standard deviations) up to 40 switching cycles.

Figure 3a depicts an archetypal metasurface consisting of a periodic array of identical GSST meta-atoms. The meta-atom dimensions are chosen such that near the telecom wavelengths, the meta-atoms support a dipole-like resonant mode (Fig. 3b inset) in their amorphous state and a quadrupole-like mode in the crystalline state. Evident from the simulated spectra in Fig. 3b, switching between the two states creates a large optical contrast at 1.49 μm wavelength, where efficient coupling to the quadrupole-like mode in the crystalline state occurs while coupling to the dipole-like mode in the amorphous state is suppressed due to optical phase mismatch.

The design principle is validated by our experimental data in Fig. 3c showing the reflectance spectra of the device after 40 crystallization-amorphization switching cycles. Complete reversible switching of the GSST material is also verified using micro-Raman measurements (Supplementary Section II). The device boasts a large absolute optical reflectance ($\Delta R$) contrast of 40% at 1.49 μm wavelength and a relative reflectance modulation ($\Delta R/R$) up to 400% at 1.43 μm wavelength. The optical contrast is substantially larger than those achieved in active metasurfaces relying on thermo-optic or electro-optic effects[14,15] and benefits from the colossal index modification of PCMs upon phase transition. The device also exhibits consistent switching characteristics with a small average cycle-to-cycle reflectance variation of 3.5% (Fig. 3c). Further details about the metasurface design and characterization are elaborated in Supplementary Section II.

**Quasi-continuous multi-state tuning**

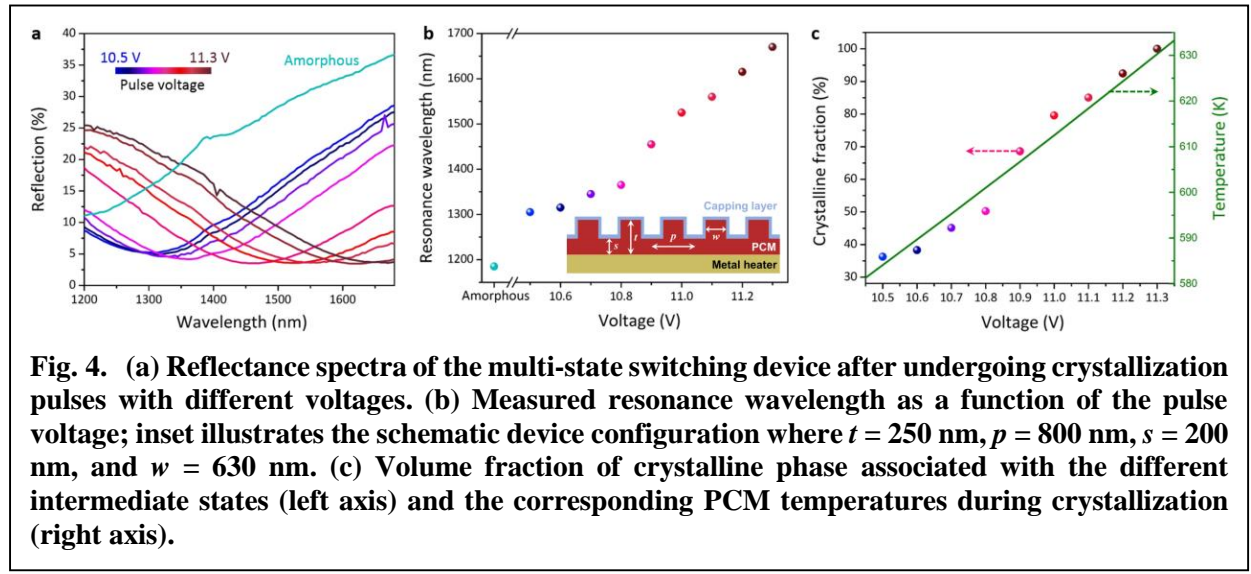

Fig. 4. (a) Reflectance spectra of the multi-state switching device after undergoing crystallization pulses with different voltages. (b) Measured resonance wavelength as a function of the pulse voltage; inset illustrates the schematic device configuration where $t$ = 250 nm, $p$ = 800 nm, $s$ = 200 nm, and $w$ = 630 nm. (c) Volume fraction of crystalline phase associated with the different intermediate states (left axis) and the corresponding PCM temperatures during crystallization (right axis).

Besides the giant index modulation, PCMs are also known for their capacity for multi-state or multi-bit operation[44–47]. As discussed in the previous section, the excellent thermal uniformity inside the meta-atoms implies that the intermediate states contain interspersed crystalline and amorphous phases of varying fractions. Precise control of the phase composition of the intermediate states can be realized through either varying the crystallization electrical pulse duration, or the pulse voltage and thereby regulating the crystallization kinetics. Here we choose the latter mechanism to demonstrate quasi-continuous tuning of the meta-atom resonance.

The device, which comprises meta-atoms arranged in a square lattice (Fig. 4b inset), is designed to attain ultra-broadband resonance tuning. Unlike devices reported by others where the PCM layer is limited to an ultrathin (< 100 nm) form factor to facilitate rapid quenching during re-amorphization and hence switching reversibility[22–25], the GSST layer in our device has a total thickness of 250 nm. Our prior work has proven that GSST exhibits improved amorphous phase stability compared to the classical GST alloys[42], which enables a much larger PCM thickness while maintaining fully reversible switching capability. The increased PCM volume affords significantly enhanced light confinement and interaction with the PCM layer, which underlies an exceptionally large resonance tuning range. Figure 4a displays reflectance spectra of the device measured after multiple crystallization cycles, where varying pulse voltages were used to access different intermediate states of the PCM. Ultra-broadband tuning of the optical resonance over half an octave, from 1190 nm to 1680 nm, was measured (Fig. 4b). This is, to the best of our knowledge, by far the largest spectral span that a non-mechanical active meta-optical device covers.

The result also allows quantitative evaluation of the GSST meta-atom crystallization kinetics. This is accomplished by extracting the crystalline phase fraction for each intermediate state via an effective medium theory (Fig. 4c left axis), and correlating the pulse voltage with corresponding meta-atom temperature using finite-element modeling (Fig. 4c right axis). Our device platform therefore not only validates multi-state switching capabilities of the PCM metasurfaces but also provides a versatile platform for quantitatively assessing the crystallization kinetics of PCMs (Supplementary Section III).

**Beam steering using active phase-gradient Huygens' surface**

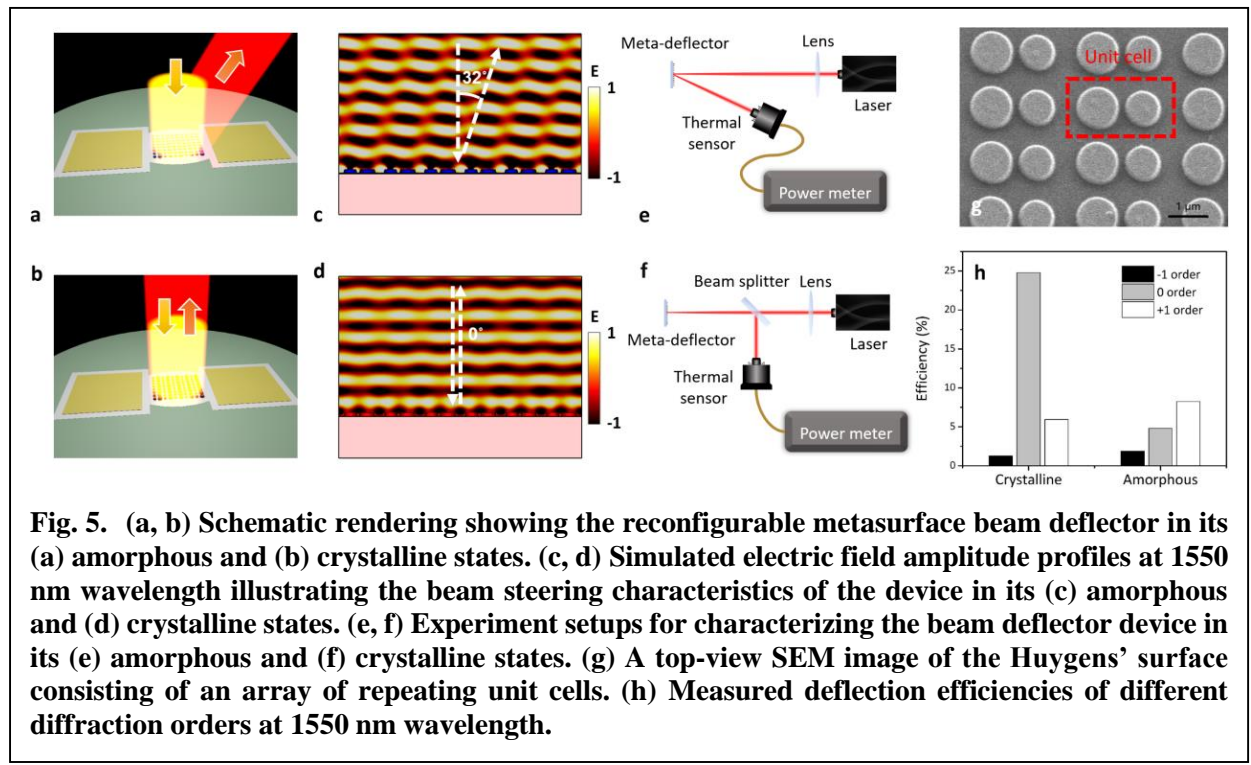

**Fig. 5. (a, b) Schematic rendering showing the reconfigurable metasurface beam deflector in its (a) amorphous and (b) crystalline states. (c, d) Simulated electric field amplitude profiles at 1550 nm wavelength illustrating the beam steering characteristics of the device in its (c) amorphous and (d) crystalline states. (e, f) Experiment setups for characterizing the beam deflector device in its (e) amorphous and (f) crystalline states. (g) A top-view SEM image of the Huygens' surface consisting of an array of repeating unit cells. (h) Measured deflection efficiencies of different diffraction orders at 1550 nm wavelength.**

Here we further show that besides amplitude or resonance tuning, the PCM-based active metasurface platform also enables facile control of optical phase and wave front by demonstrating a polarization-insensitive reconfigurable metasurface beam deflector. The device functions as a Huygens' surface[52,53] consisting of only two cylindrical elements (as shown in Fig. 5g). Details of the aggressively discretized Huygens' metasurface design[54] are furnished in Supplementary Section IV. Figures 5c and 5d present simulated field patterns of the devices in its amorphous and crystalline states, respectively. The device is designed such that for both transverse electric (TE) and transverse magnetic (TM) polarized light at 1550 nm wavelength normally incident on the device, the reflected beam couples to the +1 mode in the amorphous state with a deflection angle of 32° and to the 0$^{th}$ order mode in the crystalline state. The polarization-independent response was experimentally validated, confirming that the TE and TM deflected beams exhibit identical intensities within our measurement uncertainty. In the crystalline (amorphous) state, the measured deflection efficiency into the $0^{th}$ (+1) order is 24.8% (8.3%). The switching contrast ratio, defined as $(I_c^0 I_a^1)/(I_c^1 I_a^0)$ where $I$ represents the light intensity, the superscripts stand for the diffraction order, and the subscripts denote the device state, was measured to be 8.6 dB, which is lower than the simulated value of 14.0 dB albeit on par with state-of-the-art active metasurface deflectors[13]. We anticipate that improved fabrication accuracy as well as adoption of advanced meta-atom designs[31,32,55] will further boost the device performance.

**Conclusion**

In this work, we have demonstrated an on-chip electrical switching platform enabling both binary switching and quasi-continuous tuning of PCM-based active metasurfaces. Compared to thermal or optical actuation schemes, electrical addressing of active metasurfaces represents an important step forward toward realizing fully-integrated, chip-scale reconfigurable optics. In addition, we show that using a new PCM, namely GSST, imparts important advantages including mitigation of optical losses and increased PCM reversible switching volume. Leveraging the enhanced light-PCM interactions, we achieved unprecedented ultra-broadband tuning of meta-atom resonances across half an octave. This exceptionally large modulation capability is critical to realizing active metasurfaces for on-demand phase and amplitude control. As an example of phase and wave front control, we further realized active switching of a Huygens' surface beam deflector using the platform. Moreover, the electrical switching design and the geometric optimization approach for large-aperture metasurfaces demonstrated herein are also transferrable to other emerging heater materials such as silicon[56,57] and graphene[50] to enable a diverse array of reconfigurable metasurface devices.

**Methods**

**Material synthesis.** GSST thin films were prepared using thermal evaporation from a single $Ge_2Sb_2Se_4Te_1$ source. Bulk starting material was synthesized using a standard melt quench technique from high-purity (99.999%) raw elements[58]. The film deposition was performed using a custom-designed system (PVD Products, Inc.) following previously established protocols[42,59,60]. Stoichiometries of the films were confirmed using wavelength-dispersive spectroscopy (WDS) on a JEOL JXA-8200 SuperProbe Electron Probe Microanalyzer (EPMA) to be within 2% (atomic fraction) deviation from target compositions.

**Device fabrication and packaging.** All the devices were fabricated on silicon wafers with 3 μm thermal oxide from Silicon Quest International. The 50 nm Ti/20 nm Pt heaters and the 10 nm Ti/100 nm Au contact pads were deposited via electron beam evaporation and patterned using lift-off. 10 nm of $Al_2O_3$ was then coated using atomic layer deposition (ALD) to prevent direct contact between GSST and the heater. The GSST metasurfaces were then deposited and patterned using poly (methyl methacrylate) (PMMA) as the electron beam lift-off resist and subsequently capped with 15 nm $Al_2O_3$ deposited using ALD. All the patterning steps were carried out using electron beam lithography (EBL) on an Elionix ELS-F125 system. The devices were subsequently wire bonded (using 0.8 mil 99% Al-1% Si wires) and mounted onto a custom-designed printed circuit board (PCB), which allowed for reproducible electrical contact (as compared to using contact probes).

**Optical simulation.** For the multi-state metasurfaces, finite-difference time-domain (FDTD) simulations were performed using the commercial software package Lumerical FDTD. Periodic boundary condition (PBC) was set along both *x* and *y* in-plane directions and perfectly matched layer (PML) boundary condition was used along the out-of-plane *z* direction. Reflectance spectra were recorded with an *x-y* plane monitor on top of the device. For the beam deflector devices, S-parameter and far-field radiation pattern calculations were carried out with the commercial full-wave computation package CST Microwave Studio using the frequency solver under unit cell boundary conditions. Optical constants for the metal heater materials and both states of GSST characterized using spectroscopic ellipsometry were employed in the simulations.

**Thermal Simulation.** The thermal simulations were performed using a 3-D finite element method (FEM) model constructed in COMSOL Multiphysics. A COMSOL built-in module (Electric Currents) was used for solving the electrical current distribution, and the Heat Transfer in Solids module was implemented for simulating the heating transfer and temperature distribution. The two modules were coupled through the Electromagnetic Heat Source model. In the Heat Transfer module, infinite element domains were adopted for the side and bottom boundaries. Convective heat flux boundary condition was used for the top surface with a heat transfer coefficient of 10 $W/(m^2 \cdot K)$. The thermal properties of GSST in the model were defined based on experimentally measured data, which we will report in detail in a separate publication. Specifically, the thermal conductivity values of GSST in the model were 0.2 $W/(m \cdot K)$ for the amorphous state and 0.4 $W/(m \cdot K)$ for the crystalline state, and heat capacity values of GSST were taken as 1.45 $MJ/(m^3 \cdot K)$ for the amorphous state and 1.85 $MJ/(m^3 \cdot K)$ for the crystalline state.

**Tunable metasurface characterization.** The metasurface devices with 200 μm square apertures have electrical resistances of approximately 20 Ω. To amorphize the device, a single 20 V, 5 μs pulse was applied. For crystallization, a single 500 ms pulse with a voltage of approximately 10 V was applied. A Renishaw Invia Reflex micro-Raman system was used for collecting Raman spectra on the devices. A Thermo Fisher FTIR6700 Fourier Transform Infrared (FTIR) spectrometer with an attached microscope was used to obtain the reflectance spectra of the devices. The reflectance spectra were calibrated using a standard gold mirror.

In the quasi-continuous tuning experiment, the device reflectance spectra were taken in-situ using a super continuum laser (NKT Photonics SuperK Extreme Continuum Laser) with a tunable notch filter (SuperK Select Multi-Line Tunable Filter) across the 1.2-1.7 μm spectral range, with the reflected light intensity quantified using an InGaAs short-wave infrared (SWIR) camera (Xenics Xeva 320 Series SWIR Camera). The reflectance spectra were similarly calibrated using a gold mirror.

**Beam deflector characterization.** The measurement setup was schematically illustrated in Figs. 5e and 5f. The device under test (DUT) was mounted on a 2-D translational stage for optical alignment. A laser source (LUNA Technologies OVA-5000) whose wavelength was fixed at 1550 nm in conjunction with a fiber-optic collimator was used for characterizing the metasurface deflector. The collimated beam was focused by a lens with a 15 cm focal length onto the device. The laser beam diameter exiting from the collimator is approximately 3 mm, thereby giving rise to an effective numerical aperture (NA) of the beam of approximately 0.01. When the weakly focused beam impinged on the DUT, the spot size was 0.14 mm, which ensures that the entire spot falls on the aperture of the DUT. An infrared camera was first used to image the diffracted spots before quantifying their intensity values using a calibrated photodetector. To measure the zero-order power, a beam splitter was inserted into the optical path at 45 degrees with respect to the optical axis. The same measurement procedure was followed as above. The total power of the laser and the transmission and reflection coefficients of all the optical components were measured to calculate the absolute reflectance of the zero order.

**Acknowledgments**

This work was funded by Defense Advanced Research Projects Agency Defense Sciences Office (DSO) Program: EXTREME Optics and Imaging (EXTREME) under Agreement No. HR00111720029. The authors also acknowledge characterization facility support provided by the Materials Research Laboratory at Massachusetts Institute of Technology (MIT), as well as fabrication facility support by the Microsystems Technology Laboratories at MIT and Harvard University Center for Nanoscale Systems. The views, opinions and/or findings expressed are those of the authors and should not be interpreted as representing the official views or policies of the Department of Defense or the U.S. Government.

**Author contributions**

Y.Z. performed material deposition, device design, and metasurface characterization. Y.Z. and J.L. fabricated the metasurfaces and conducted device characterization. C.F. conceived and designed the Huygens' surface. B.A., M.Y.S., S.D. and C.R. assisted in device characterization. S.A. helped with device modeling. J.B.C., C.M.R., and V.L. measured the multi-state metasurfaces. M.K. prepared the bulk materials. T.G., C.R.B., K.R., H.Z., and J.H. supervised the research. All authors contributed to technical discussions and writing the paper.

**Competing financial interests**

The authors declare no competing financial interests.